\documentclass[runningheads,a4paper,oribibl]{llncs}
\usepackage{mathptmx}
\usepackage{amssymb}
\usepackage{graphicx}
\usepackage{enumitem}
\usepackage{subfigure}
\usepackage{multirow}
\usepackage[table,xcdraw]{xcolor}

\usepackage{url}
\newcommand{\keywords}[1]{\par\addvspace\baselineskip
\noindent\keywordname\enspace\ignorespaces#1}

\begin{document}

\mainmatter  

\title{The Role of Compliance in Heterogeneous\\Interacting Agents: Data from Observations}

\titlerunning{The Role of Compliance in Heterogeneous Interacting Agents}

\author{Stefania Bandini\textsuperscript{1, 2} \and Luca Crociani\textsuperscript{1} \and Giuseppe Vizzari\textsuperscript{1} \and \\Flavio Soares Correa da Silva\textsuperscript{3} \and Andrea Gorrini\textsuperscript{1}}

\authorrunning{Bandini, S., Crociani, L., Vizzari, G., Correa da Silva, F., Gorrini, A.}

\institute{
\textsuperscript{1} Complex Systems and Artificial Intelligence research center, 
Univ. Milano - Bicocca (Italy) \\
\textsuperscript{2} Research Center for Advance Science and Technology, The University of Tokyo (Japan)\\
\textsuperscript{3} Institute of Mathematics and Statistics, University of S\~ao Paulo (Brazil)
}

\toctitle{Lecture Notes in Computer Science}
\tocauthor{Authors' Instructions}
\maketitle

\begin{abstract}
The dynamics of agent-based systems provide a framework to face the complexity of pedestrian/vehicle interactions in future cities, in which the compliance to traffic norms plays a fundamental role. The data of an observation performed at a non-signalized intersection are presented to provide useful insights for supporting the future development of agent-based models. Results focus on drivers\rq\ compliance to crossing pedestrians, describing potentially conflictual interactions among heterogeneous agents. The discussion closes with the potential applications of the collected data set for modelling the phenomenon.
\end{abstract}


\section{Introduction and Related Work}
\label{sec:intro}

Agent-based modelling and simulations of pedestrian and crowd dynamics have been increasingly reported in the technical and scientific specialized literature. Scientific communities started to incorporate agent-based systems to improve the expressiveness of traditional approaches and to simulate the complex behaviour of people and traffic in outdoor and indoor urban scenarios. The intrinsically dynamical properties of agent-based models offer a research framework to face the complexity of the future cities~\cite{masthoff2007agent}, offering new possibilities to incorporate and integrate the growing presence of autonomous entities/artefacts both physical (e.g. autonomous vehicles) and virtual (e.g. data coming from heterogeneous sources: social media, distributed sensors etc.).

The development of agent-based models and systems requires to check the quality, robustness and plausibility of the obtained simulations against real data, in order to tackle decision making problems related to urban mobility. Recent literature contains a wide range of methods and study cases supporting this view~\cite{campanella2014quantitative}. 
The aim of this paper is to present a real case of data collection performed to collect useful insights about pedestrian-vehicles interactions at non-signalized intersections, supporting the future development of a heterogeneous agent-based system to simulate the phenomenon.

From pioneering works, several models have been developed and applied for the simulation of pedestrian and vehicular dynamics, including both CA and particles models~\cite{duives2013state}. These two approaches have, separately and independently, produced a significant impact, yet efforts characterized by an integrated model considering the simultaneous presence of vehicles and pedestrians are not as frequent or advanced. With the notable exception of~\cite{Helbing2005}, most efforts in this direction are relatively simplistic, narrow (i.e. targeting extremely specific situations), homogeneous for the simulated entities, and they are often not validated against real data~\cite{Godara2007,Zeng2014}. In this framework, we consider the possibility to model and simulate the complex aspect of drivers' compliance to pedestrian yielding rules in the context of interactions among heterogeneous agents, and its implications on self-organization dynamics.

\section{Observation Results}\label{sec:observation}

\begin{figure}[t]
\begin{center}
\includegraphics[width=.75\textwidth]{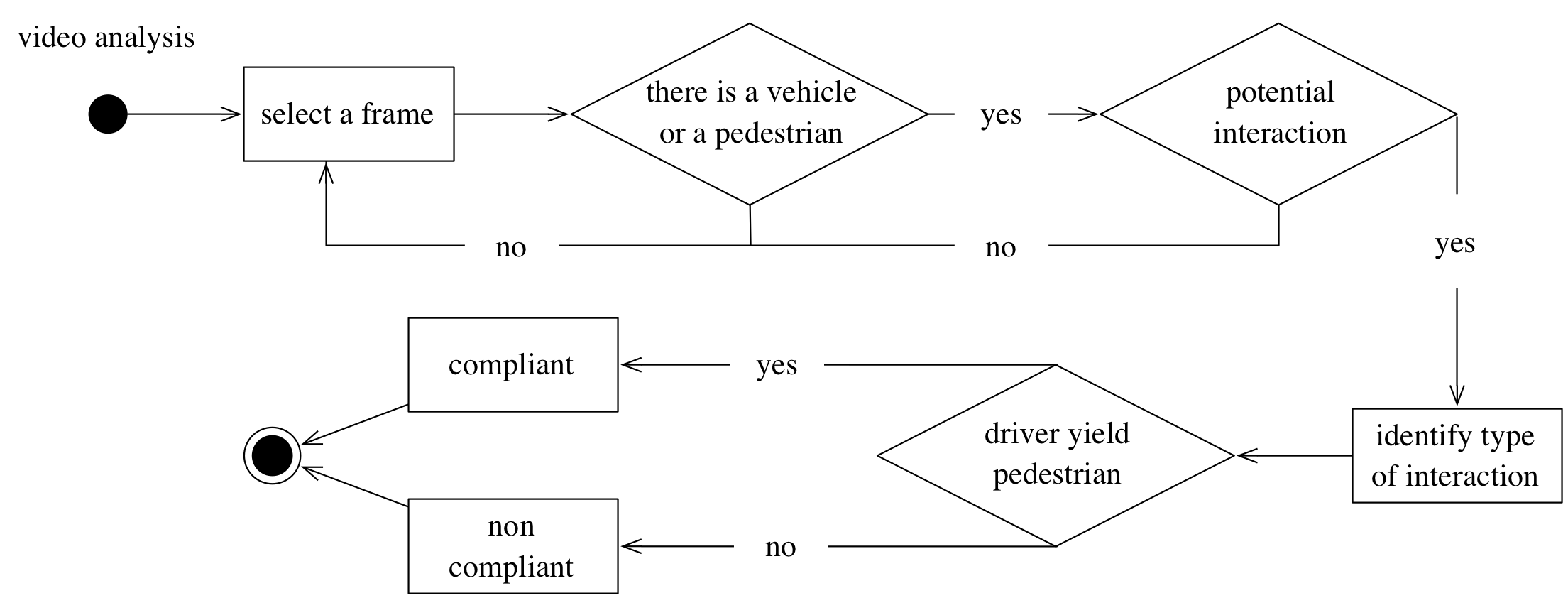}
\caption{The work flow for selecting the sample of crossing episodes from the video images.}
\label{fig:if}
\end{center}
\end{figure}

A video recorded observation~\cite{PED2016crossing} has been performed at a non-signalized intersection in Milan (from noon, 73 minutes), characterized by a high number of pedestrian accidents in the past years. We selected from the video a sample of 812 crossing episodes considering only the cases in which one vehicle directly interacted with one or more pedestrians (see Fig.~\ref{fig:if}). At non-signalized intersections traffic laws require drivers to yield those pedestrians who are already occupying the zebra, but also those who are localized nearby or in correspondence of the curb waiting to cross. The level of compliance of drivers to pedestrians\rq\ right-of-way have been estimated considering the position of crossing pedestrians with respect to the curb (i.e. pedestrian about 1.5 meter far from the curb, waiting at the curb, crossing on the zebra) and to the direction of travel of vehicles (i.e. pedestrian from the near or the far side-walk).  

Preliminary analyses on results (see Tab. \ref{tab:compliance}) showed that 48\% of the total number of crossing episodes was characterized by non-compliant drivers with crossing pedestrians from the two side-walks. A multiple linear regression was calculated to predict the percentage of non-compliant drivers per minute based on: (\emph{i}) number of vehicles per minute (18.89 veh/min in average; p = 0.007, significant predictor) and (\emph{ii}) number of crossing pedestrian per minute (8.01 ped/min in average; p~$<$~0.001, significant predictor). A significant regression equation was found [F(2,70) = 14.526, p~$<$~0.001], with R$^2$ of 0.293. This demonstrates that the non-compliance of drivers is negatively determined by traffic conditions and positively determined by pedestrian flows on zebra. Despite the low level of drivers\rq\ compliance, no accidents or risky situations have been observed, thanks to the self-organization of the system based on pedestrians\rq\ yielding/collaborative behaviour to approaching cars.

\vspace{-3mm}
\begin{table}[t]
\small
\centering
\caption{Results about the drivers\rq\ compliance to the right-of-way of crossing pedestrians.}%\vspace{0.5cm}
\label{tab:compliance}
\begin{tabular}{|l|c|c|}
\hline
\textbf{Types of pedestrian/vehicle interaction} & \textbf{\begin{tabular}[c]{@{}c@{}} Compliant \end{tabular}} & \textbf{\begin{tabular}[c]{@{}c@{}} Non-compliant \end{tabular}} \\ \hline
\begin{tabular}[c]{@{}l@{}}Ped. approaching/waiting/crossing from the near side-walk  \end{tabular}& 191 (46.14\%) & 223 (53.86\%) \\ \hline
\begin{tabular}[c]{@{}l@{}}Ped. approaching/waiting/crossing from the far side-walk  \end{tabular}& 230 (57.69\%) & 168 (42.21\%) \\ \hline
\end{tabular}
\end{table}

\begin{figure}[h!]
\begin{center}
\includegraphics[width=.4\textwidth]{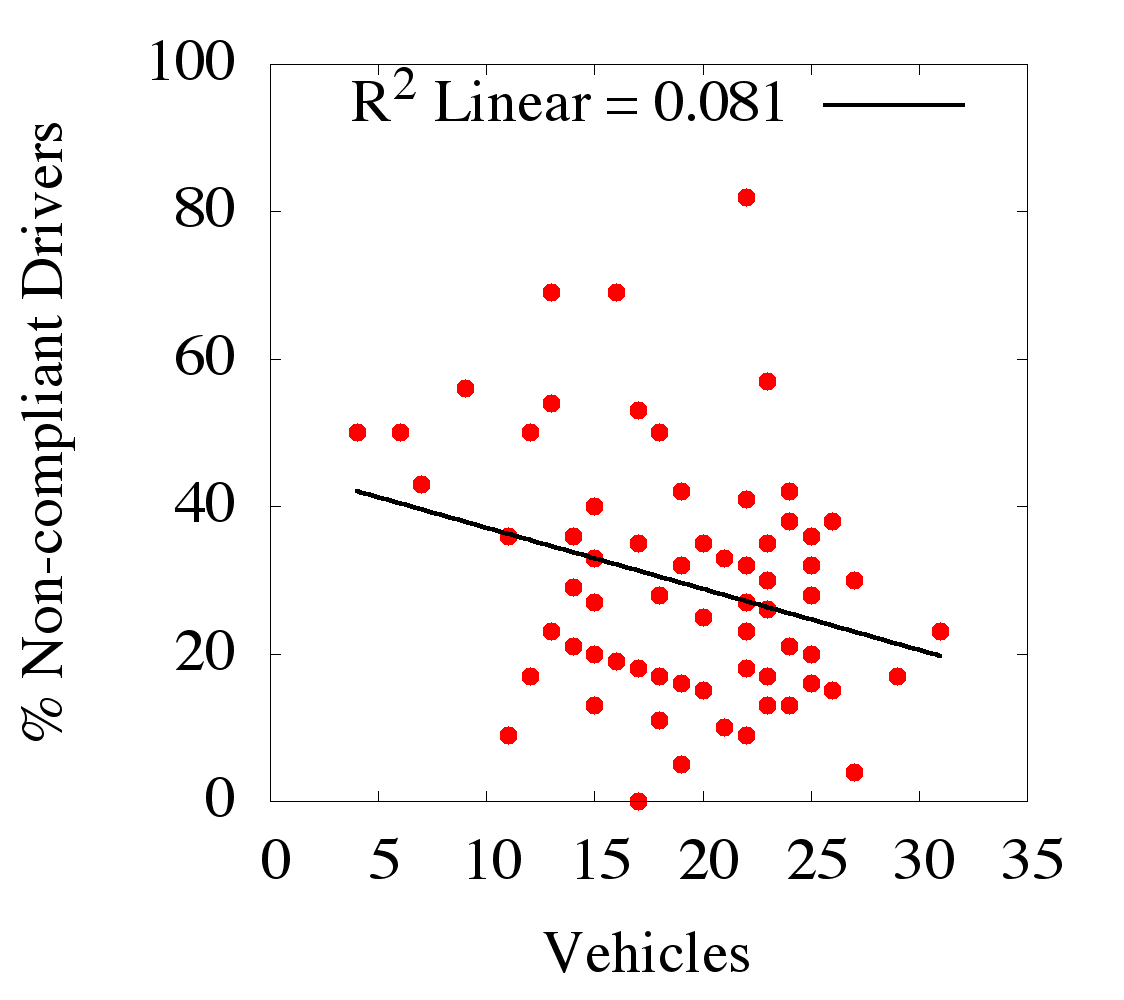}
\includegraphics[width=.4\textwidth]{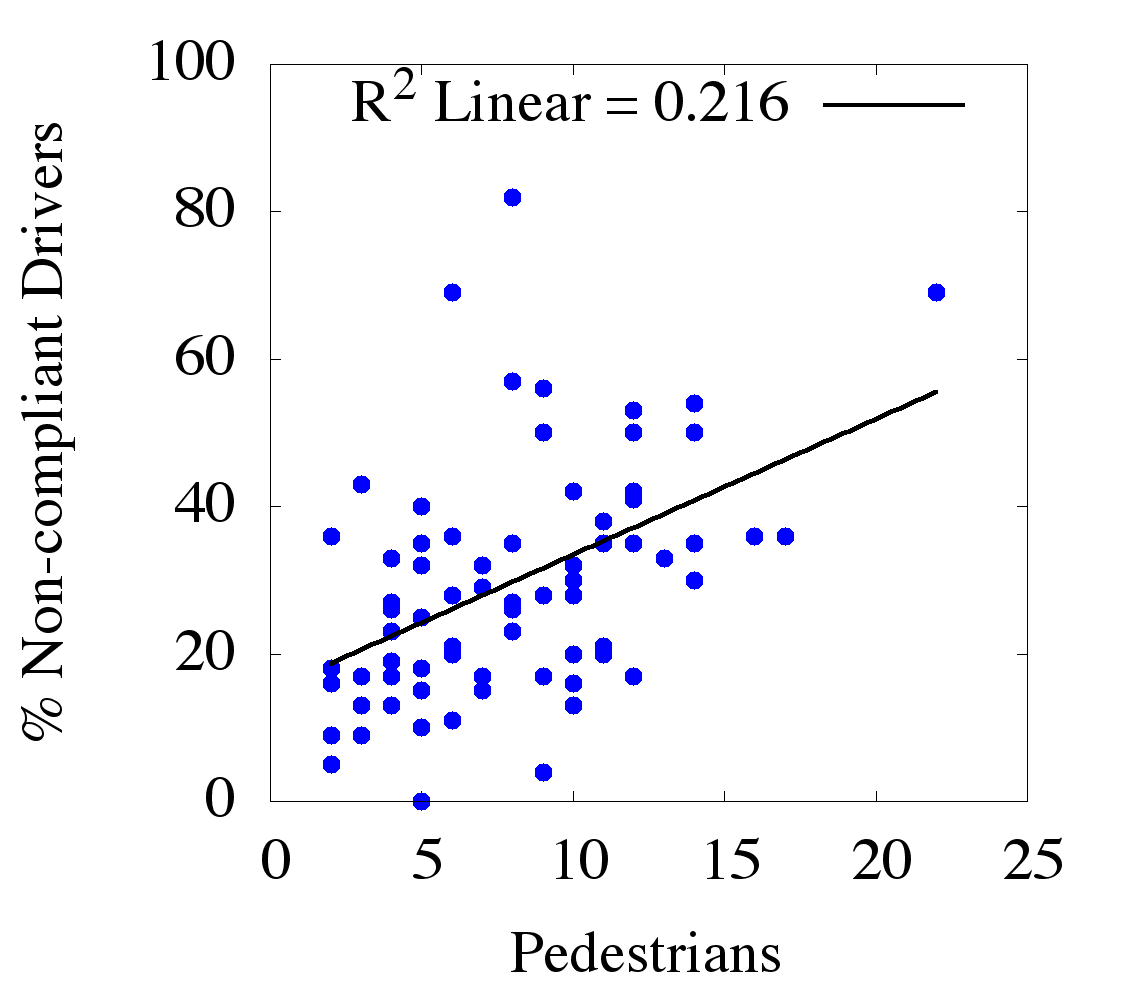}
\caption{The relation between non-compliant drivers and the number of vehicles (left) and crossing pedestrians (right) in the observed intersection. Each data refers to one minute of the video.}
\label{fig:data}
\end{center}
\end{figure}

\section{Discussion}\label{sec:model}
\begin{figure}[t]
\centering
\centering
\subfigure[]{\includegraphics[width=.8\textwidth]{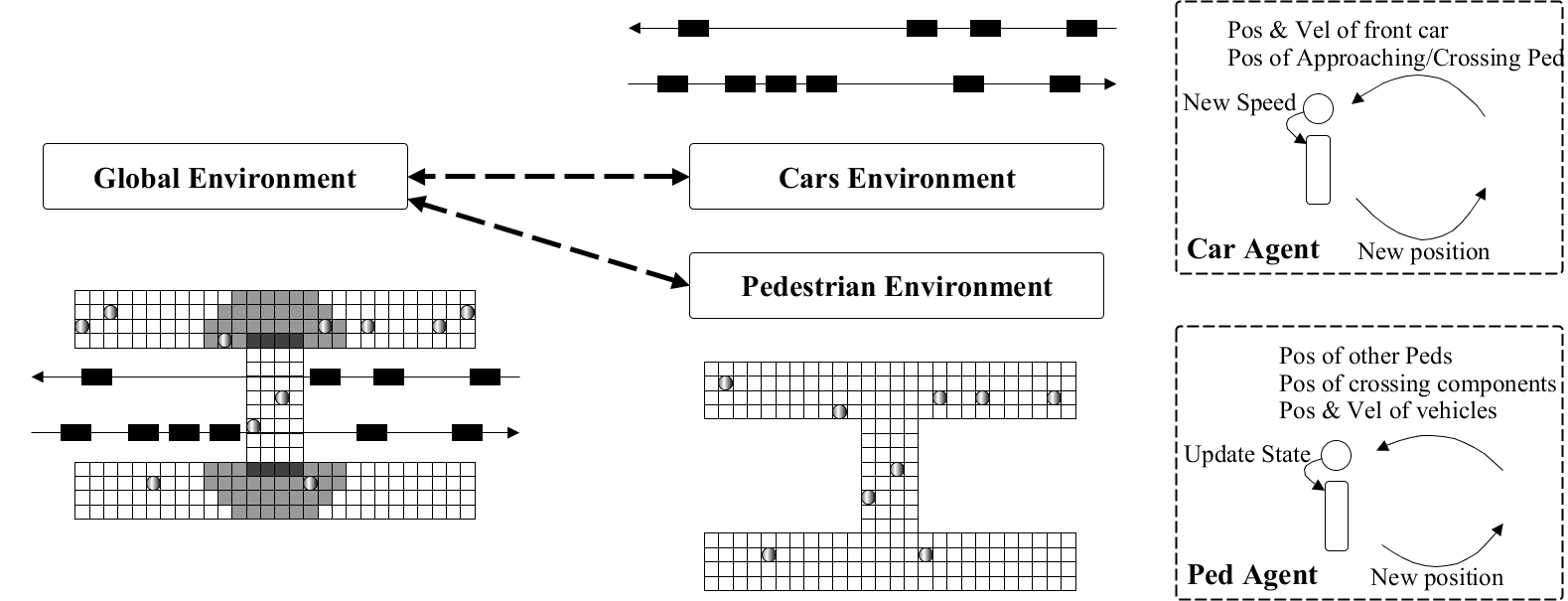}}
\subfigure[]{\includegraphics[width=.8\textwidth]{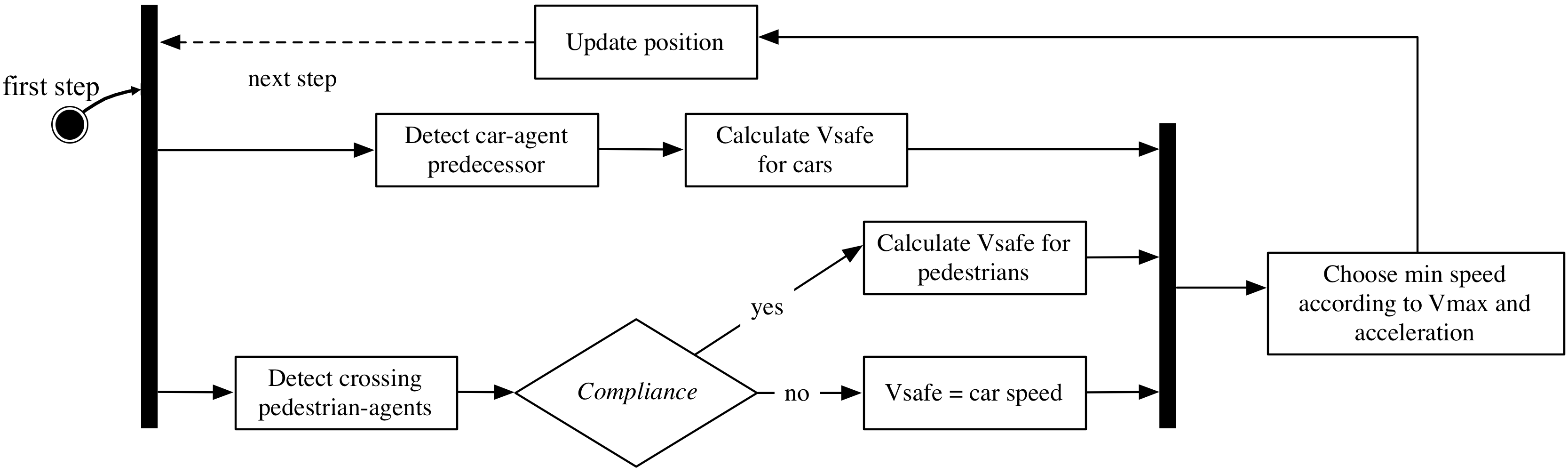}}
\caption{A description of the environment and the two classes of agents (a). Car-agents life-cycle considering their possible non-compliance with respect to the presence of pedestrian-agents (b).}
\label{fig:schema}
\end{figure}

The results showed in the previous section represent applicable insights towards the extension of a model for the analysis of pedestrian crossings~\cite{CrocianiVizzari2014}, considering potentially conflictual interactions among heterogeneous agents. As shown in Figure~\ref{fig:schema}(a), the model is based on the  integration of two independent models for the simulation of vehicles, moving in continuous lanes, and pedestrians, moving in a 2-dimensional discrete environment. The two environments are superimposed, and car-agents perceive pedestrian-agents while they are crossing or in the nearby of the curb (grey cells in Fig.~\ref{fig:schema}(a)) and vice-versa. The interactions between them are described in Figure ~\ref{fig:schema}(b). Pedestrian-agents consider the speed and distance of cars to avoid collisions, giving way to non-compliant vehicles. The compliance of car-agents is mainly influenced by the necessary braking distance. On the other hand, according to a fixed probability that will be set on the collected data on compliance, car-agents can deliberately avoid to stop even if the braking distance is sufficient, requiring  pedestrian-agents to yield. 

The potential applications of such research are related to the possibility to test the effect of non-compliance on emergent observables like: near accident situations, exposure to accidents, traffic capacity of the road and Level of Service~\cite{milazzo1999quality}. In addition, this kind of research is also potentially relevant to complement studies on autonomous vehicles~\cite{10.1371/journal.pone.0149607} to evaluate future transportation scenarios in Smart Cities. 

% \bibliographystyle{splncs03}
% \bibliography{template} % name your BibTeX data base

\end{document}